\documentclass[aps,prl,epsfig,onecolumn]{revtex4}

\usepackage[dvips]{graphics}
\usepackage{graphicx}
\usepackage{amsfonts}
\usepackage{amssymb}
\usepackage{amsmath}

\begin{document}

\title{Geometric stabilization of extended $S=2$ vortices in
two-dimensional photonic lattices:
theoretical analysis, numerical computation and experimental results}
\author{K.J.H. Law$^1$, D. Song$^2$, P.G. Kevrekidis$^1$, J. Xu$^2$,
and Zhigang Chen$^{2,3}$}
\affiliation{
$^1$ Department of Mathematics and
Statistics, University of Massachusetts,
Amherst MA 01003-4515, USA \\
$^{2}$ TEDA Applied Physics School, Nankai University, Tianjin, 300457,
China \\
$^3$ Department of Physics and Astronomy, San Francisco State
University, San Francisco, CA 94132 USA
}

\begin{abstract}
In this work, we focus our studies on the subject of nonlinear
discrete
self-trapping of $S=2$ (doubly-charged) vortices in two-dimensional
photonic
lattices, including theoretical analysis, numerical computation and
experimental demonstration. We revisit earlier findings about $S=2$
vortices with a discrete model, and find that $S=2$ vortices extended
over eight lattice sites can indeed be stable (or only weakly unstable)
under certain conditions,
not only for the cubic nonlinearity previously used, but also for a
saturable nonlinearity more relevant to our experiment with a biased
photorefractive nonlinear crystal. We then use the discrete analysis
as a
 guide towards numerically identifying stable (and unstable) vortex
solutions in a more realistic continuum model with a periodic
potential.
Finally, we present our experimental observation of
such geometrically extended
$S=2$ vortex solitons in optically induced lattices under both self-focusing
and self-defocusing nonlinearities,
and show clearly that the
$S=2$ vortex singularities are preserved during nonlinear propagation.

\end{abstract}

\maketitle

\section{Introduction}

Over the last two decades, the study of Hamiltonian lattice systems,
as well as of continuum models with periodic potentials
has been a subject of increasing interest
\cite{reviews}.
These systems arise in a  diverse host of physical contexts,
describing, e.g.,
the spatial dynamics of optical beams in coupled waveguide arrays
or photorefractive crystals or optically induced photonic
lattices in
nonlinear optics \cite{reviews1},
the temporal evolution of Bose-Einstein condensates (BECs)
in optical lattices
in soft-condensed matter physics \cite{reviews2}, or
the DNA double strand in biophysics \cite{reviews3}, among many others.

One of the main thrusts of work in these directions has been
centered around the investigation of existence and stability of
localized solitary wave solutions. In two dimensions, such
structures can be discrete solitons \cite{solit,solit_physd} or discrete
vortices (i.e., structures that have topological
charge over a discrete contour) \cite{vort,vort_physd}.
Optically-induced photonic lattices in photorefractive
crystals such as strontium barium niobate (SBN) have been used as
an ideal platform for the observation of those predicted soliton
structures.
Indeed, the theoretical proposal \cite{solit} of such lattice solitons
was followed quickly by their experimental
realization in 2D induced lattices
\cite{moti1,moti2}, subsequently leading to the
observation of a host of novel solitons
in this setting, including
 dipole \cite{dip}, multipole \cite{quad}, necklace \cite{neck},
and rotary \cite{rings} solitons as well as discrete \cite{vortex1,vortex2}
and gap \cite{motihigher}
vortices. In addition to lattice solitons, photonic lattices
have enabled observations of other intriguing phenomena such as
higher order Bloch modes \cite{neshev2},
Zener tunneling \cite{zener}, and localized modes
in honeycomb \cite{honey}, hexagonal \cite{rosberg2}
and quasi-crystalline \cite{motinature1} lattices,
and Anderson localization \cite{motinature2}
(see, e.g., the recent review \cite{moti3} for additional examples).
In parallel, experimental development in the area of
BECs closely follows, with prominent recent results
 including the observation
of bright, dark and gap solitons in quasi-one-dimensional settings
\cite{bec}, with the generation of similar structures in higher dimensions
being experimentally feasible for BECs trapped in optical
lattices \cite{ol2d,vortol1}.

Earlier experimental work on discrete vortex solitons
\cite{vortex1,vortex2}
has mainly focused on vortices of unit topological
charge (i.e., $S=1$, with a 2$\pi$ phase shift circle around a
discrete contour). However,
more recently both in optics \cite{zc_s2,dsong} and in BEC
\cite{ketts2,corns2} (so far in the absence of the lattice for the
latter case), the study of higher charge vortices
has been of interest. In particular, in the emerging
area of hexagonal \cite{rosberg2,rosberg3} and honeycomb \cite{honey}
lattices, it has been predicted \cite{ourhexhon,ournewpra} and
experimentally observed \cite{ourexppra} very recently that a
higher-order
vortex with topological charge $S=2$ is {\it more} stable
than a fundamental vortex with unit charge ($S=1$) when
self-trapped with a focusing nonlinearity.

The prototypical nonlinear dynamical lattice
associated with the above systems is the so-called discrete
nonlinear Schr{\"o}dinger equation (DNLS) \cite{pgk}.
In the context of that model,
it has been predicted that genuine $S=2$ vortices
turn out to be unstable in the case of a
square lattice \cite{vort,vort_physd}.
This instability, however, can be avoided as proposed in two
separate ways: a more intrusive one in which the central site of the
contour is eliminated, introducing a defect therein \cite{ours2}, as
well as a less intrusive one involving
the configuration of $S=2$ vortex
on a ``rhomboidal'' contour as 8-site excitation
in the square lattices, proposed solely on the basis
of numerical observation as in \cite{mjs2}.
Our aim in the
present paper is to understand from a theoretical perspective
the {\it geometric stabilization} (as we will call it) of
$S=2$ vortices, and to illustrate its generic nature in discrete
systems, by considering the case of optically induced
photonic lattices with a photorefractive nonlinearity
(where there are some special intricacies of
the theoretical analysis that we also illustrate).
Based on our theoretical analysis of the discrete model, we then
consider a more realistic continuum model of beam propagation with
a periodic potential, illustrating
that the $S=2$ vortices geometrically extended to eight sites can
indeed be linearly stable even in the
continuum model. Finally, we corroborate these theoretical
and numerical analyses with direct experimental observation of
self-trapped $S=2$ vortices whose topological charges are sustained
during propagation throughout the nonlinear medium, which can be clearly
contradistinguished from prior observations of breakup or charge-flipping
of such high-order vortices [28].

The structure of our presentation is as follows. In
section II we examine the model of the discrete nonlinear
Schr{\"o}dinger type with both Kerr and saturable nonlinearities, where
our analytical findings are compared with numerical
bifurcation results. In section III we consider the continuum
model of beam propagation in photorefractive media with a periodic
potential relevant to our experiment for the
focusing and briefly also for the defocusing case. In section IV we present
our experimental results. Section V concludes
the paper with a number of interesting directions proposed for
future studies.

\section{Discrete Models: Analysis and Numerics}

The DNLS with both Kerr and saturable nonlinearities can be written as:
\begin{eqnarray}
i \dot{u}_{m,n}= -\varepsilon \Delta_2 u_{m,n} -
\mathcal{N}(|u_{m,n}|^2) u_{m,n}
\label{eqn1}
\end{eqnarray}
where $\mathcal{N}(|u|^2)=|u|^2$ for the typical DNLS with Kerr
nonlinearity, while
$\mathcal{N}(|u|^2)=-1/(1+|u|^2)$ in the saturable one (notice
that both models share the same small amplitude limit).
In the above, the overdot denotes the derivative with respect
to the evolution variable, while $\Delta_2 u_{m,n}= u_{m+1,n}+u_{m-1,n}
+ u_{m,n+1} + u_{m,n-1} - 4 u_{m,n}$ stands for the discrete Laplacian.
We seek sationary solutions of the form $u_{m,n}=\exp(i \mu t) v_{m,n}$
[notice that we denote the evolution variable as $t$, although in
optics it represents the propagation distance $z$], obtaining
the steady state equation:
\begin{equation}
[\mu-\varepsilon \Delta_2 - \mathcal{N}(|v_{m,n}|^2)] v_{m,n}=0.
\label{sta_eq}
\end{equation}

Our considerations will start at the so-called anti-continuum (AC)
limit of $\varepsilon=0$, where it is straightforward to solve
the steady state equations for a given $\mu$. In the DNLS case,
each site can be $v_{m,n} = \sqrt{\mu} \exp(i \theta_{m,n})$ or
$v_{m,n}=0$ and in the saturable case, we have
$v_{m,n} = \sqrt{-1/\mu-1} \exp(i \theta_{m,n})$ or
$v_{m,n}=0$. In order to consider cases where the solutions
have the same amplitude between our two examples, we will
select, without loss of generality $\mu=1$ for the DNLS case,
while $\mu=-1/2$ for the saturable one.

It is worthwhile to notice that in the AC limit, solutions
can be obtained with arbitrary phase profiles. However, out of
all the possible profiles, it is relevant to examine which
ones may survive for $\varepsilon \neq 0$. To do so, we
expand the solution into a power series e.g., $v=v_0 + \varepsilon v_1
+ \dots$, and write down the solution of the system order by order.
To leading order the relevant equation for $v_1$ will yield
\begin{center}
\begin{math}\label{discrete_lin2}
\mathcal{J}(v_0,\varepsilon=0) \bordermatrix{& \cr & v_1 \cr & v_1^* \cr} -
\bordermatrix{& \cr & \Delta_2  & 0\cr
 & 0 & \Delta_2 \cr}
 \bordermatrix{& \cr & v_0 \cr & v_0^* \cr} =0.
\end{math}
\end{center}

where $\mathcal{J}$ is the Jacobian of Eqs. (\ref{sta_eq}) [with
respect to the variables $v_0$ and $v_0^*$] evaluated
at the solution for $\varepsilon=0$:

\begin{equation}
\label{energy} {\cal J}(v,\varepsilon) = \left( \begin{array}{cc} 1 - 2 \partial({\cal N}v)/\partial{v} &
- \partial({\cal N}v)/\partial{v^*} \\ - \partial({\cal N}v)^*/\partial{v} & 1 - 2
\partial({\cal N}v)^*/\partial{v^*} \end{array} \right) - \varepsilon
\bordermatrix{& \cr & \Delta_2  & 0\cr
 & 0 & \Delta_2 \cr} .
\end{equation}

For each of the excited
sites (indicated by $j$)
there corresponds a
zero eigenvalue of
the Jacobian with
an eigenvector which 
vanishes away from
the $j$-th block.
Projecting to this eigenvector, as explained in \cite{solit_physd,vort_physd}
yields the Lyapunov-Schmidt conditions for the persistence of
the solution.

\begin{figure}[t]
\includegraphics[width=8cm,height=6cm,angle=0,clip]{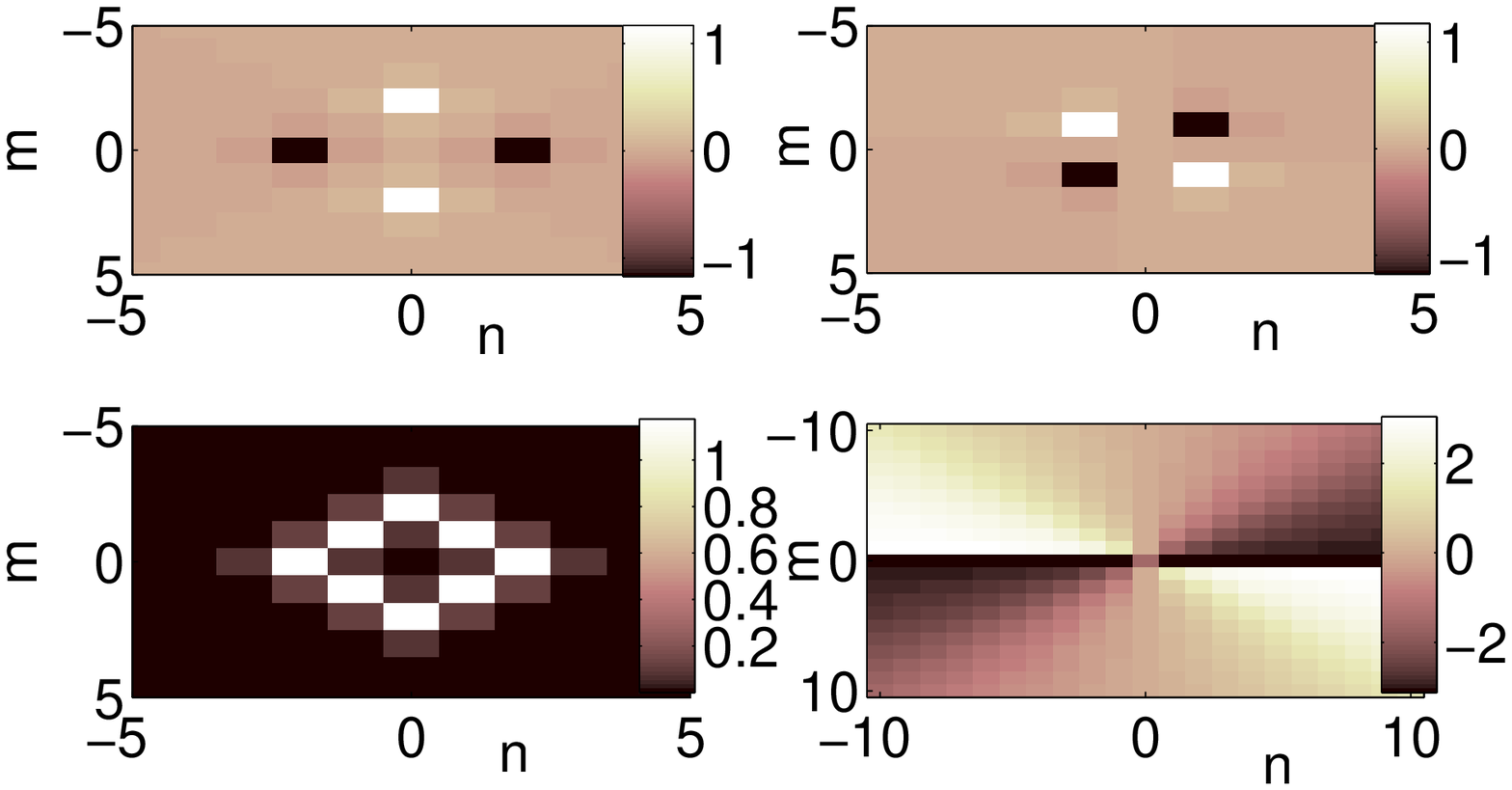}
\includegraphics[width=8cm,height=6cm,angle=0,clip]{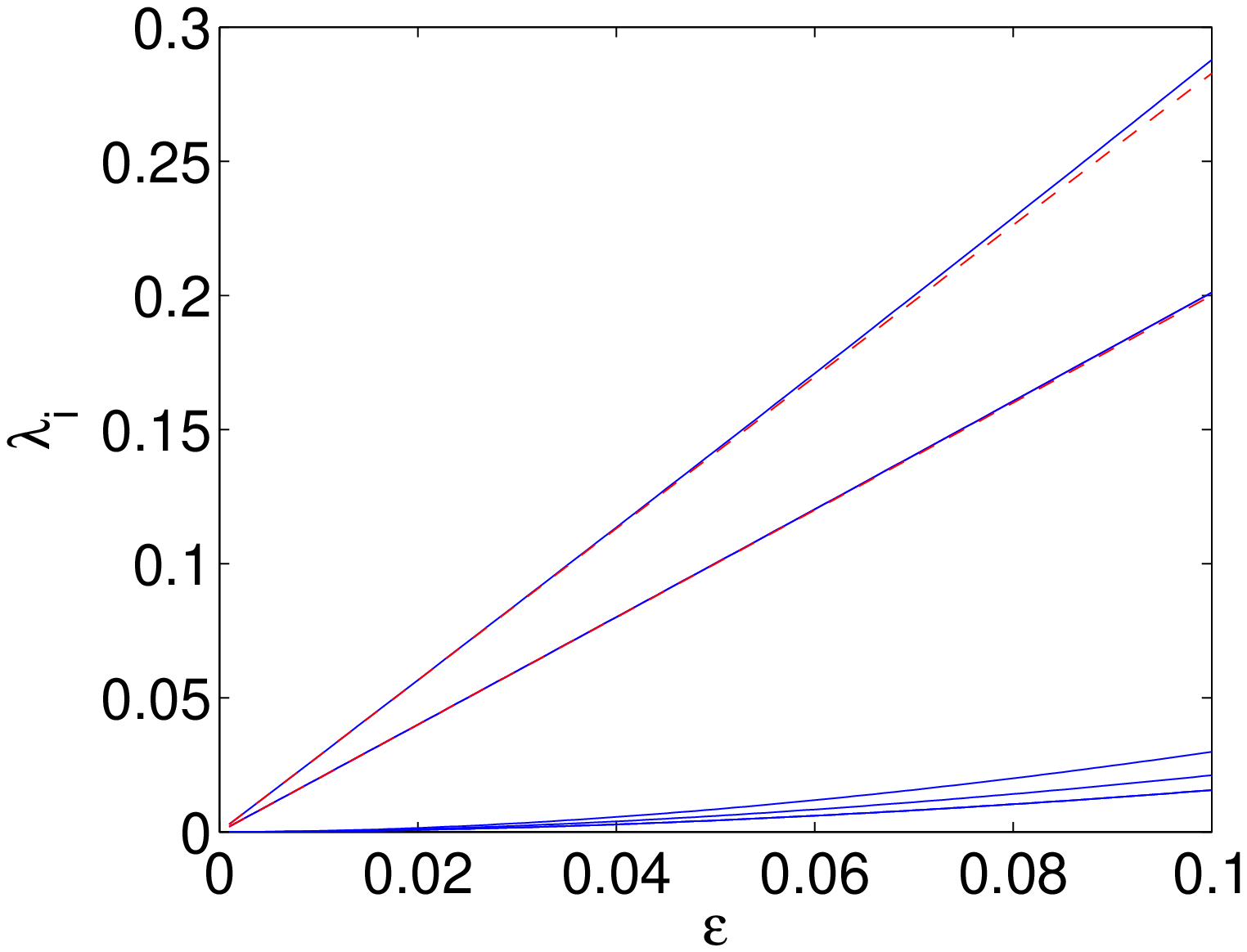}
\caption{(Color online)
The left panels show a typical example (for $\varepsilon=0.1$)
of the geometrically stabilized
rhomboidal $S=2$ vortex obtained from DNLS with Kerr (cubic) nonlinearity
(top left is the real part, top right is its imaginary part, bottom
left shows the amplitude and bottom right the phase distribution).
The right panel shows the dependence of the corresponding eigenvalues
as a function of $\varepsilon$. The solid lines show the full numerical
results for the imaginary parts of the eigenvalues $\lambda_i$
as a function of the coupling constant $\varepsilon$, while the two dashed
lines (the lowest one of which is indistinguishable from the corresponding
solid line) show the explicit analytical predictions in the text, which
are in very good agreement with the corresponding numerics.}
\label{bfig1}
\end{figure}

First, let us clarify the notation we will use in this paper.
When the principal axes of a square lattice are normally oriented in
horizontal and vertical directions,
we  name the structure a {\it square vortex} when it
excites horizontal and vertical (nearest-neighbor)
sites within a square contour, and
likewise, a {\it rhomboidal vortex} when it excites diagonal
(next-nearest-neighbor)
sites in a rhomboidal contour (see Fig. \ref{bfig1}).
We shall stick to such notation later for the continuum model
even when the principal axes are diagonally oriented, as typically
used in experiments with induced lattices \cite{vortex1,vortex2,zc_s2}.
For the case of the $S=2$ square vortex, different conditions
have been analyzed in \cite{vort_physd,ours2} and will not
be presented here.
Instead, we present here the case of the geometrically stabilized $S=2$
rhomboidal vortex. In the latter case, if we denote the vortex by
indices $1, \dots, 8$, starting from the top of the rhombus,
then it is straightforward to see that the fundamental difference
between the square and rhomboidal settings is that
{\it all} interactions between adjacent sites in the rhombus
are {\it next nearest neighbor}
ones, and hence they come in not at the leading order in perturbative
corrections (i.e., at O$(\varepsilon)$), but rather at O$(\varepsilon^2)$,
having the following form:
\begin{eqnarray}
g_{2 j+1} &=& 2 \sin(\theta_{2 j +1}-\theta_{2j}) + 2 \sin(\theta_{2j+1}-
\theta_{2 j +2})
\label{ex1}
\\
g_{2 j} &=& 2 \sin(\theta_{2 j}-\theta_{2 j-1}) + 2 \sin(\theta_{2 j}-
\theta_{2 j+1}) + \sin(\theta_{2 j}-\theta_{2 j-2}) + \sin(\theta_{2 j}-\theta_{2 j+2})
\label{ex2}
\end{eqnarray}
where $j=1,2,3,4$, with periodic boundary conditions (i.e., site
$0$ corresponds to site $8$, site $9$ to site $1$ etc.).
Both in the DNLS, as well as in the saturable lattice
case, the $S=2$ solution with $\Delta \theta= \pi/2$ among
adjacent sites (so that an accumulation of phase of $8 \times \pi/2=4 \pi$
occurs around the discrete contour for these
extended 8-site vortices) naturally satisfies
the  above persistence conditions.

However, the more crucial question is that of stability of the
pertinent solutions. As initially illustrated in \cite{solit_physd,vort_physd},
perhaps not surprisingly, the matrix that bears the relevant
information is the {\it Jacobian} of the persistence conditions
 $({\cal M})_{j,k} \equiv \partial g_j/\partial \theta_k$.
To leading order, in the cubic case,
the eigenvalues $\gamma_j$ of this Jacobian
are associated with the small eigenvalues of the full problem
via
\begin{eqnarray}
\lambda_j^2 = 2 \varepsilon \gamma_j.
\label{eigs}
\end{eqnarray}

\begin{figure}[t]
\includegraphics[width=8cm,height=6cm,angle=0,clip]{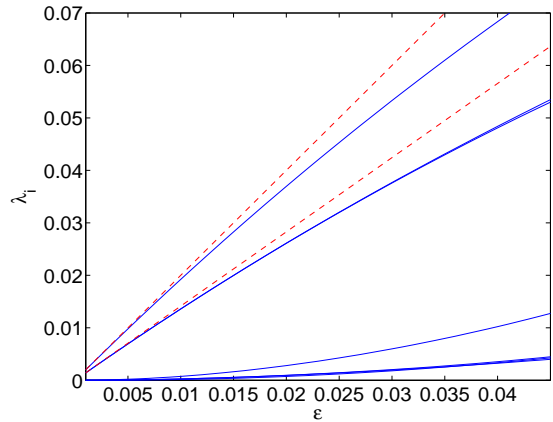}
\caption{(Color online) Plots of the imaginary part of the eigenvalues similarly to
  the
right panel of Fig. \ref{bfig1}, but obtained from DNLS with saturable
nonlinearity.}
\label{bfig2}
\end{figure}

The linear stability problem of solutions of
Eq. (\ref{eqn1})
is then determined from the eigenvalue problem:
\begin{equation}
\label{eigenvalue}
{\cal J}(v,\varepsilon) \mbox{\boldmath $\psi$} = i \lambda
{\cal \sigma} \mbox{\boldmath $\psi$}.
\end{equation}
and ${\cal \sigma}$ is
$$
\left( \begin{array}{cc} I & 0 \\ 0 & -I \end{array} \right).
$$
Note that this system can equivalently be rearranged into $2$-by-$2$
block formation.
The discrete vortex is called spectrally unstable if there exist
$\lambda$ and $\mbox{\boldmath $\psi$}$ in
the problem (\ref{eigenvalue}), such that ${\rm Re}(\lambda) > 0$.
Otherwise, the discrete vortex is called spectrally stable.

It is straightforward to observe that in the limit of
$\varepsilon=0$, only the excited sites yield a set of $N$
(i.e., as many as the number of such sites) pairs of null
eigenvalues, while the non-excited sites yield eigenvalues
with $\lambda=\pm 1$ (which will ``become'' the continuous
spectrum of the problem); in the saturable case,
this part of the spectrum is at $\pm 1/2$.
Out of these $N$ vanishing eigenvalue pairs
in the AC limit, only one can be sustained at the origin for $\varepsilon
\neq 0$, due to the preserved phase invariance of the solution,
while the remaining ones have to move off of the origin,
as dictated by Eq. (\ref{eigs}). Interestingly, in the case
of the unstable square contour vortex, the phase change of
$\pi/2$ across sites renders zero all elements of the corresponding
tridiagonal (in that case) Jacobian of the persistence conditions
which, there, read $g_j = \sin(\theta_{j}-\theta_{j+1}) +
\sin(\theta_j - \theta_{j-1})$. This is the so-called
``super-symmetric'' case of \cite{vort_physd} which needs to be
treated at a higher order and, as a result, leads to higher
order eigenvalues \cite{vort_physd,ours2}.
These studies illustrated that in addition to a pair at $0$
and another pair of higher order, there is a quadruple pair
$\lambda = \pm \sqrt{2} \varepsilon i$, a single pair $\pm \sqrt{\sqrt{80}+8}
\varepsilon i$ and a {\it real} pair (responsible for the instability)
$\lambda= \pm \sqrt{\sqrt{80}-8} \varepsilon$. The investigation of
these works illustrated that this real pair was due to the interaction
between the 4 sites adjacent to the central one of the vortex
(to 2nd order, as mediated by the central site of the vortex).
That is why the defect-induced stabilization worked in \cite{ours2}
via its exclusion of this type of interactions.

However, stabilization also ensues, in a geometric fashion,
in the case of the rhomboidal $S=2$ pattern. What happens in this
case is that indeed the odd sites of the vortex still interact
between them through the central site, however, now this interaction
is geometrically ``screened'' by their lower order
interaction with the even-numbered sites within the contour.
As a result the instability is no longer mediated.
More specifically, the analytical calculation in the case
of the DNLS yields an $8 \times 8$ Jacobian, which for
$\Delta \theta=\pi/2$ has 4 zero eigenvalues (which will
become nonzero at a higher order), while the remaining
4 eigenvalues satisfy
\begin{eqnarray}
\gamma_j= -4 \sin^2 \left(\frac{j \pi}{4}\right),
\label{eigs1}
\end{eqnarray}
yielding, in addition to the phase-invariance induced persistent
zero pair, $\gamma_{1,3}=-2$ and $\gamma_2=-4$.
Because the interaction arises due to 2nd order neighbors,
as shown in \cite{solit_physd,vort_physd}, the relevant
contribution is $\lambda_j=\pm \sqrt{2 \gamma_j} \varepsilon$,
yielding a double pair $\lambda_{1,3}=\pm 2 \varepsilon i$
and a single pair $\lambda_2=\pm 2 \sqrt{2} \varepsilon i$,
in excellent agreement with the numerical results, as shown
in Fig. \ref{bfig1} (especially, the $\pm 2 \varepsilon i$
prediction can not be distinguished from its numerical counterpart).
Notice that, additionally to the above O$(\varepsilon)$ eigenvalue pairs,
there are also five pairs
of smaller eigenvalues, a double and two single ones,
as well as one at $0$, due to the persistence of the gauge
symmetry.

 In the saturable case, we also observe numerically the same
geometric stabilization effect, as is illustrated by the
corresponding eigenvalues, for this case, of Fig. \ref{bfig2}.
However, here there is an interesting theoretical feature that
is worth discussing. In the recent exposition of the saturable
case in \cite{rothos}, it was illustrated for nearest neighbor
contours that the formula associating the eigenvalues of the
full problem with those of the reduced Jacobian is:
$\lambda_j^2=(1/2) \gamma_j \varepsilon^2$, due to the differences in the
corresponding linearization operators. However, we observed
numerically in the present ({\it next-nearest-neighbor}) setting
that this formula no longer holds. In particular, it is
found that the dashed lines in Fig. \ref{bfig2}, which optimally
match the numerical findings for small $\varepsilon$, are given
by $\lambda=\pm 2 \varepsilon i$ and $\lambda=\pm \sqrt{2} \varepsilon i$,
namely they are less than their cubic counterparts by a factor
of $\sqrt{2}$ (rather than a factor of $2$, as per the relation of
\cite{rothos}). Retracing step-by-step, the derivation of
section 5 of \cite{solit_physd}, one can indeed theoretically
identify this nontrivial difference, when repeating the relevant
calculation for the saturable case. In particular, it is true
(we have checked that this does hold true for saturable nonlinearities
also in 1d next-nearest-neighbor contours) that generally for
saturable nonlinearities but {\it next-nearest-neighbor} contours,
the small eigenvalues of the full problem are given by:
\begin{eqnarray}
\lambda^2 {\bf c} = \varepsilon^2 {\cal M}_2 {\bf c}
\label{revised}
\end{eqnarray}
where ${\cal M}_2$ denotes the next-nearest-neighbor Jacobian.
In the cubic case, the corresponding formula (cf. (5.15) of
\cite{solit_physd})
has an extra factor of $2$ in the right hand side.
The resulting eigenvalues $\lambda_j=\pm \varepsilon \sqrt{\gamma_j}$
are compared to the full numerical results, yielding once again
good agreement for small $\varepsilon$ in Fig. \ref{bfig2}.

\begin{figure}[t]
\includegraphics[width=20cm,height=8cm,angle=0,clip]{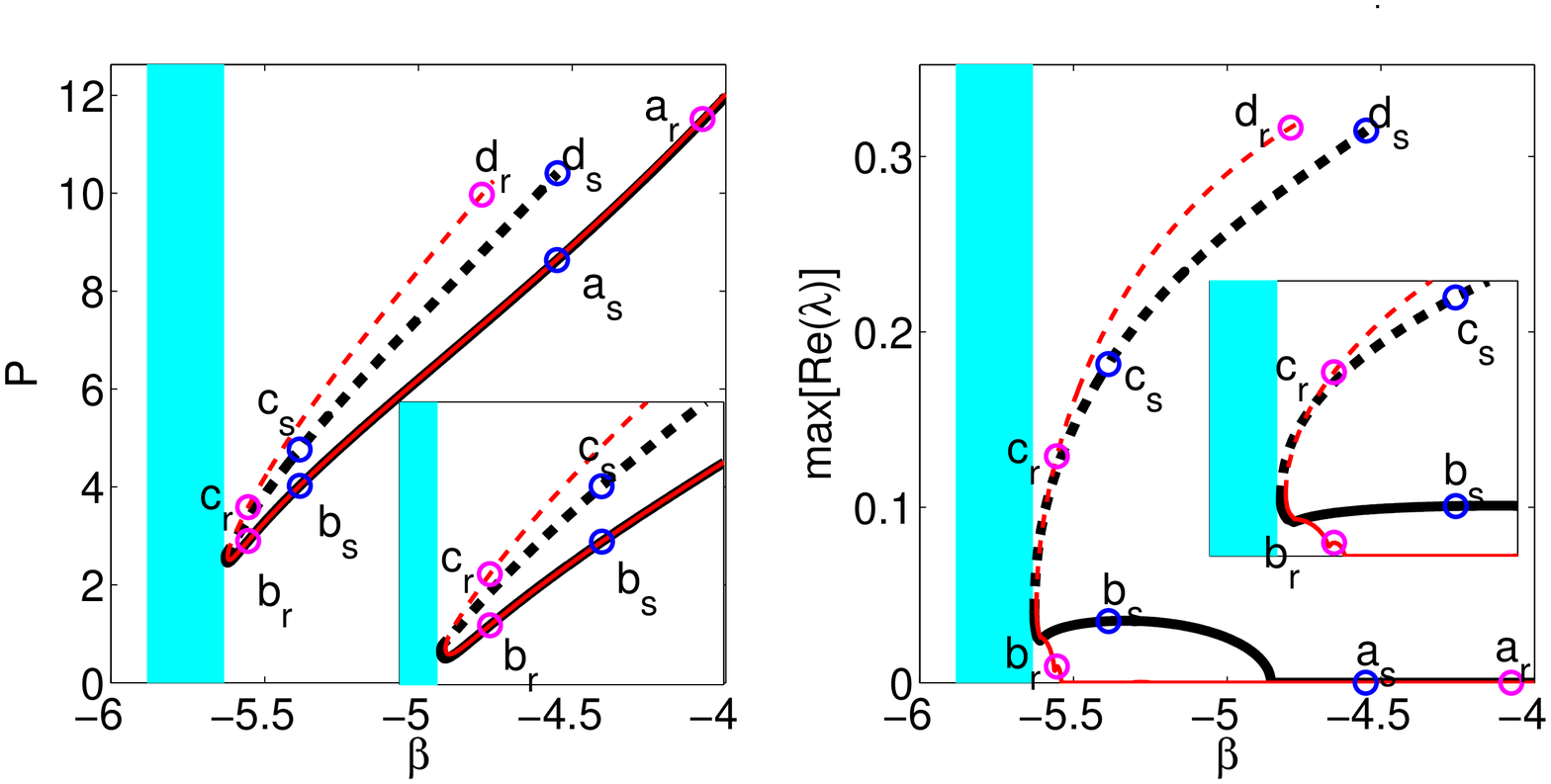}\\
\includegraphics[width=8cm,height=6cm,angle=0,clip]{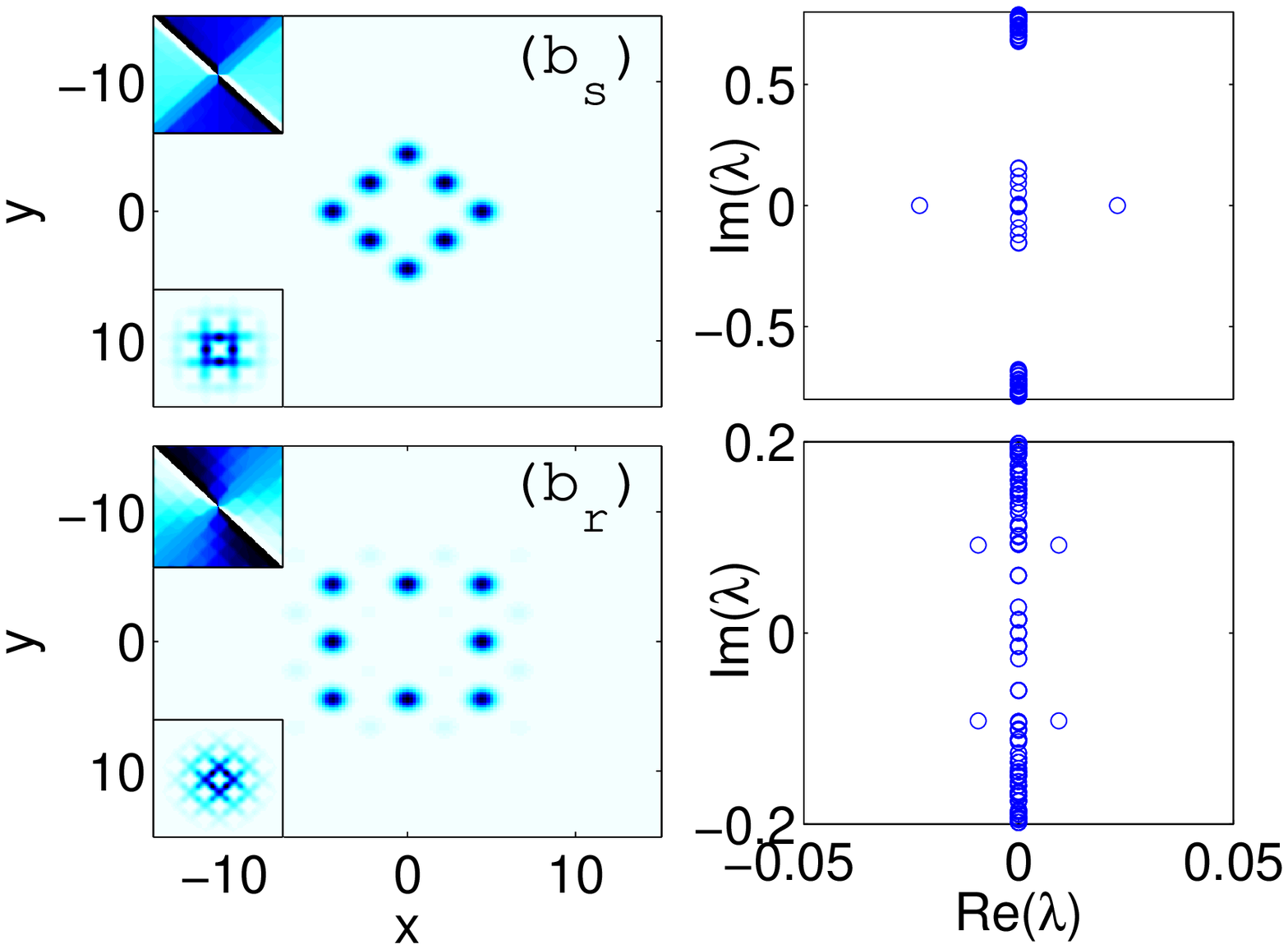}
\includegraphics[width=8cm,height=6cm,angle=0,clip]{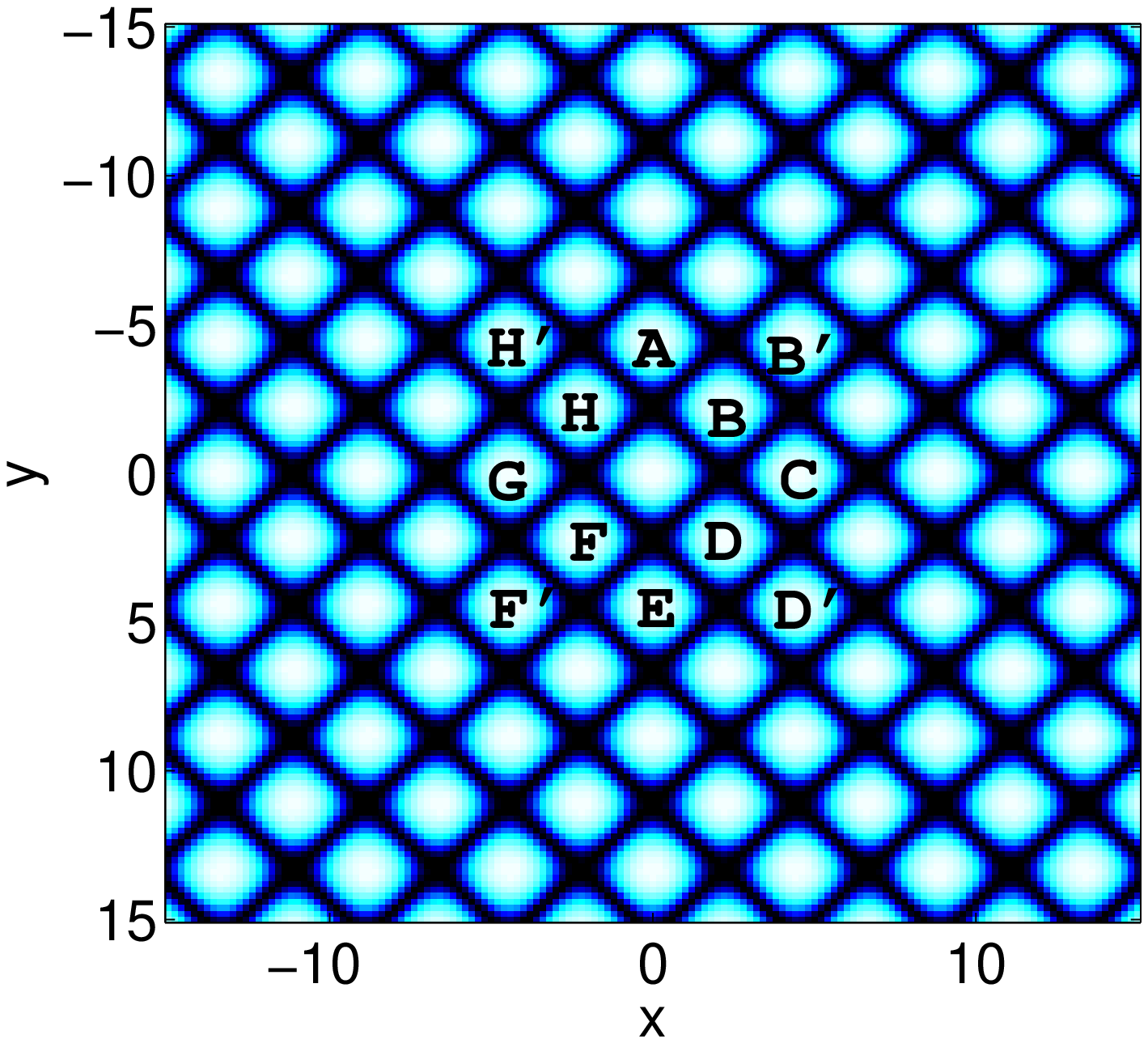}
\caption{(Color online)
The complete bifurcation structure of the $S=2$ square (thick, black)
and rhomboidal (thin, red) vortices are presented in the top panels,
with the power curve on the left and the stability curve on the right
represented by the maximum real part of the linearization spectrum.
The insets show
close-ups near the saddle-node bifurcations that occur for both branches
near the band edge.  The bottom right panel is the lattice intensity
pattern, where the eight vortex sites are marked by letters. The
``square'' vortex, as described in the discrete setting, is indicated
by
A-H, while the ``rhomboidal''vortex is indicated by AB'CD'EF'GH'.
The panels in the bottom left
show the modulus, $|U|^2$, of prototypical solutions
for the square, ($b_s$), and rhomboidal,
($b_r$ - after destabilization close to the band edge), vortices as
defined in the text.
To the right of these are their respective linearization spectra and  the
insets in the top show the complex argument, or phase, while those
to the bottom show the modulus in Fourier space (with axes in the standard
horizontal and vertical orthogonal directions).}
\label{cont_1}
\end{figure}

\begin{figure}[t]
\includegraphics[width=20cm,height=8cm,angle=0,clip]{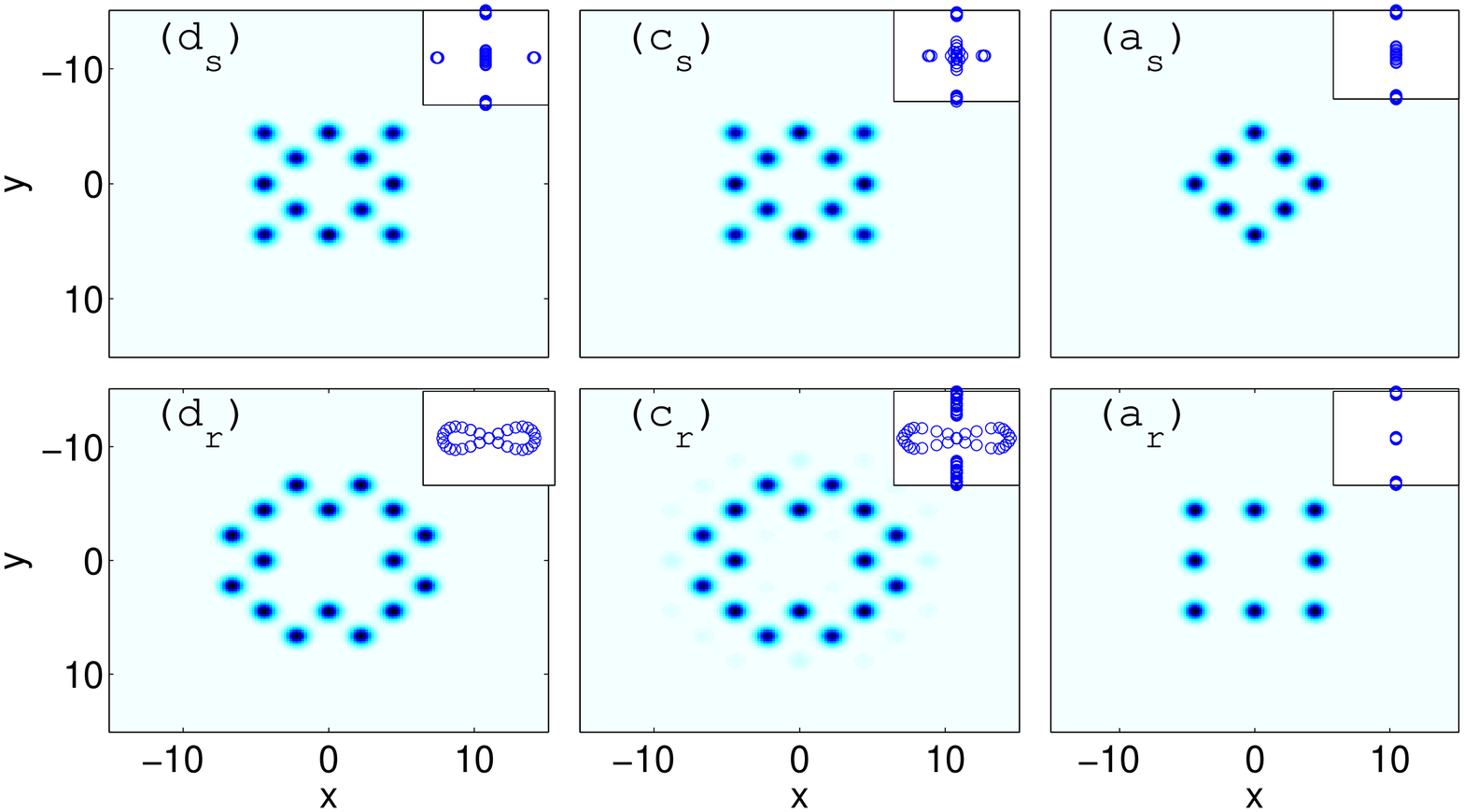}\\
\caption{(Color online) The modulus, $|U|^2$, of the remaining solutions marked in the
bifurcation diagram in the top panel of Fig. \ref{cont_1}.  The
linearization spectra are given in the insets in the top right.
All solutions are for extended s=2 vortices with similar phase
structures,
but only the 8-site excitation in the form
of square or rhomboidal vortices is stable.}
\label{cont_2}
\end{figure}

\section{Continuum Models: Bifurcation Analysis and Dynamical Evolution}

For our consideration of the continuum problem,
we use the non-dimensionalized version of the
photorefractive model with saturable nonlinearity,
as developed in detail in \cite{quad,yang04_3},
in the following form:
\begin{eqnarray}
i \dot{u} = -\Delta_2 u +
\mathcal{N}(|u|^2) u
\label{eqn1c}
\end{eqnarray}
where $\Delta_2$ is now the 2D continuum Laplacian and
$\mathcal{N}(|u|^2)=E/(1+I_{ol}+|u|^2)$.

Here, $u$ is the slowly varying amplitude of the probe beam
normalized by the dark irradiance of the crystal $I_d$, and
\begin{equation}
I_{ol}=I_0\cos^2\left(\frac{x+y}{\sqrt2}\right)\cos^2\left(\frac{x-y}{\sqrt2}\right),
\end{equation}
is a square optical lattice
intensity function in units of $I_d$. Here $I_0$ is the lattice peak
intensity, $z$ is the propagation distance (in units of
$2k_1D^2/\pi^2)$, $(x, y)$ are transverse distances (in units of
$D/\pi$), $E_0$ is the applied DC field  (in units of
$\pi^2(k^2_0n^4_eD^2r_{33})^{-1}$), $D$ is the lattice spacing, $k_0 =
2\pi/\lambda_0$ is the wavenumber of the laser in the vacuum,
$\lambda_0$ is the wavelength, $n_e$ is the  unperturbed refractive
index of the crystal for the extraordinarily polarized light,
$k_1 =k_0n_e$, and $r_{33}$ is the electro-optic
coefficient for the extraordinary polarization. In line with the
experiment,
we choose the lattice intensity $I_0 = 5$ (in units of $I_d$).
A plot of the lattice is shown in the bottom right panel of
Fig.\ \ref{cont_1}, also for
illustrative purposes regarding the locations of the lattice sites
excited by the vortex beam. Notice the lattice is now oriented
diagonally,
in accordance
with the experiment, which implies that it is rotated by
$\pi/4$
with respect to the x-y oriented lattice in the discrete model.
As such,
the rhombus in the present diagonal-oriented lattice is tantamount to the
square in the x-y oriented lattice in discrete setting, and likewise
the
square would correspond to the rhombus. Nevertheless, we still call
the
vortex extended to cover ABCDEFGH sites a {\it square vortex}, and that
extended to cover AB'CD'EF'GH' sites a {\it rhomboidal vortex},
according to the notation defined earlier.
We choose the remaining physical parameters
consistently with the experiment as
\[
D = 28\ \mu\textrm{m},\quad\lambda_0 = 0.5\ \mu\textrm{m},\quad n_e = 2.3,\quad r_{33} = 280\ \textrm{pm/V}.
\]
Thus, in the numerical results presented below,
one $x$ or $y$ unit corresponds to 8.92 $\mu$m,
one $z$ unit corresponds to 4.6 mm, and one $E$ unit corresponds to 10.17 V/mm in physical units.
We also set
$E=13.76$ which corresponds in dimensional units to $140$V/mm. %

\begin{figure}[t]
\includegraphics[width=8cm,height=10cm,angle=0,clip]{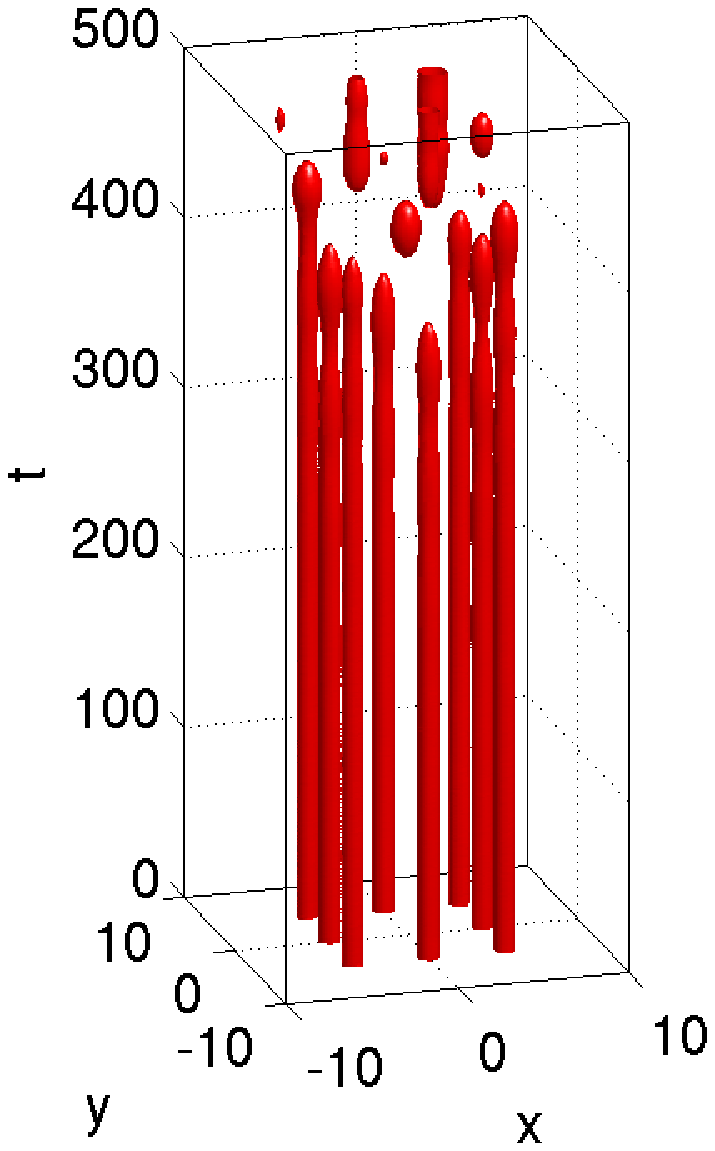}
\includegraphics[width=8cm,height=10cm,angle=0,clip]{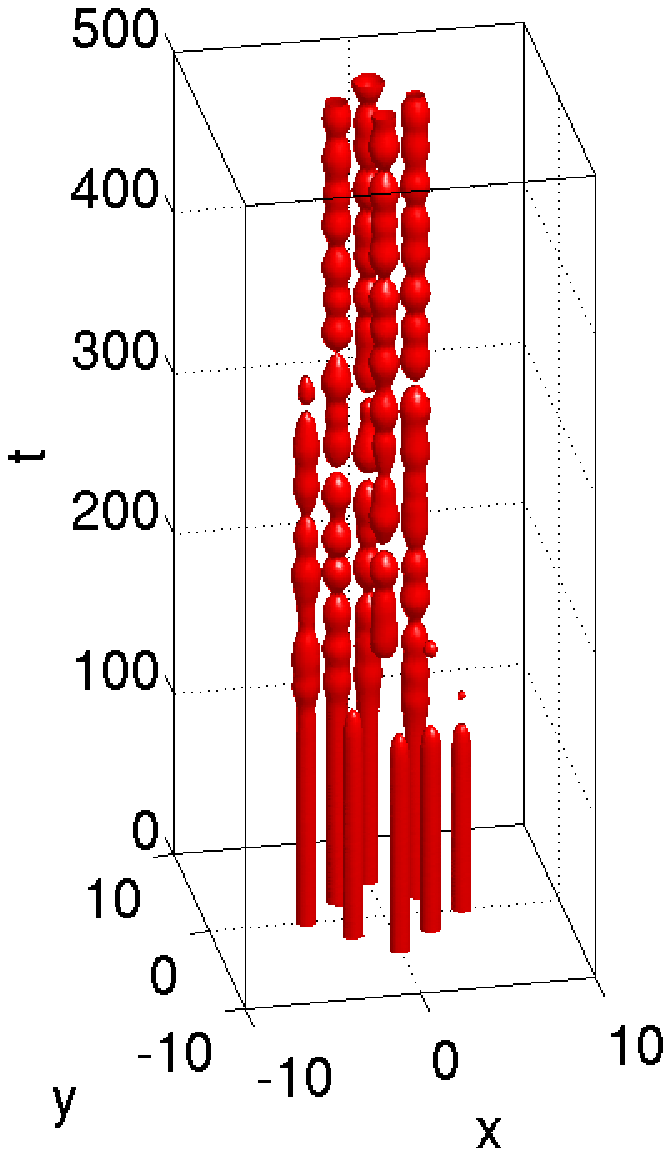}
\caption{(Color online) The modulus, $|u(t)|^2$, of the dynamical evolution of slightly
perturbed solutions from the bottom left
of Fig. \ref{cont_1} are presented above by characteristic density isosurfaces
of half-max the initial amplitude, i.e.
$D(x,y,t)=\{(x,y,t);|u(x,y,t)|^2=(1/2){\rm max}_{x,y}|u(x,y,0)|^2\}$.
The initial condition is
$u(0)=U_s(1+0.05 UN[-1,1])$, where $U_s$ is the corresponding exact solution
and $UN[-1,1]$ is a uniform random variable in
$[-1,1]$.  For the solutions of Fig. \ref{cont_1}
the left image depicts the very mild instability of the unstable
solution $b_s$ very close to the band edge, while the right image displays
the strong instability of the configuration $b_r$.}
\label{cont_3}
\end{figure}

It is well known that this model admits solutions of the form
$u(x,y,t) = U(x,y) e^{i\beta t}$ for propagation constants in
the semi-infinite gap, computed here as $\beta>-5.632$.  We
compute continuations of solutions in this region, as shown in the top
panels of Fig. \ref{cont_1} for both the square (thick, black)
and geometrically stabilized rhomboidal (thin, red) S=2
configurations \footnote{All steady state numerical computations
are performed on a square $30 \times 30$ grid with $\Delta x=0.2$
(which also mildly affects the band edge), so the computational grid
is $151 \times 151$.  The number of periods is about $7$ in each
direction and sufficient for our localized states, which occupy
2 periods in each direction, to decay well before the boundary.}.
The rhomboidal vortex (with square contour in the present
diagonal lattice) is clearly
stable through most of the interval, in consonance with our
calculations
in the discrete problem.
On the other hand, it is somewhat
surprising to note that, for $\beta$ sufficiently far from the linear
spectrum, the square vortex
(with rhomboidal contour in present diagonal lattice)
can also be stable.
There is
clearly a bifurcation of the expected
real eigenvalue pair through
the origin, however, at $\beta \approx -4.85$.  This confirms the
prediction of the discrete model, although at the same time
it indicates that its results should not be expected to be uniformly
valid within the semi-infinite gap.
The prototypical vortex solutions $(b_s, b_r)$ are presented
in the bottom left panels of Fig. \ref{cont_1}, along with phases and
corresponding Fourier space profiles (insets), as well as
linearization spectra (right).  The remaining solutions
associated with the continuation of the relevant branches are
presented together with linearization spectra insets in Fig. \ref{cont_2}.
One can observe that similarly to what was shown for single-charge
vortices in \cite{jianke_rec}, the more strongly unstable
counterparts of square and rhomboidal structures consist of
patterns occupying additional wells of the periodic potential.

The dynamics of perturbed solutions from the bottom left
of Fig. \ \ref{cont_1} are presented in Fig. \ref{cont_3}
by characteristic density
isosurfaces
of half-max the initial amplitude, i.e.
$D(x,y,t)=\{(x,y,t);|u(x,y,t)|^2=(1/2){\rm max}_{x,y}|u(x,y,0)|^2\}$
\footnote{The computational grid for the dynamical evolution is 16
periods in each direction, using $\Delta x=0.3$,
leading to a grid of $237 \times 237$.  The periods are matched exactly
with periodic boundary conditions since these large structures with
high angular momentum radiate to the boundary with small amplitude
even for this large domain.}.
The left image depicts the very mild instability of the unstable
configuration denoted above as  $b_s$
very close to the band edge, while the right image displays
the strong instability through a real eigenvalue of the configuration
denoted by $b_r$.
The mild instability of the former does not manifest itself
until after $t=300$, while the strong instability of the latter
manifests itself by $t=100$.  Also, note that a four site
breathing structure appears to persist in the latter case.

\begin{figure}[t]
\includegraphics[width=8cm,height=7cm,angle=0,clip]{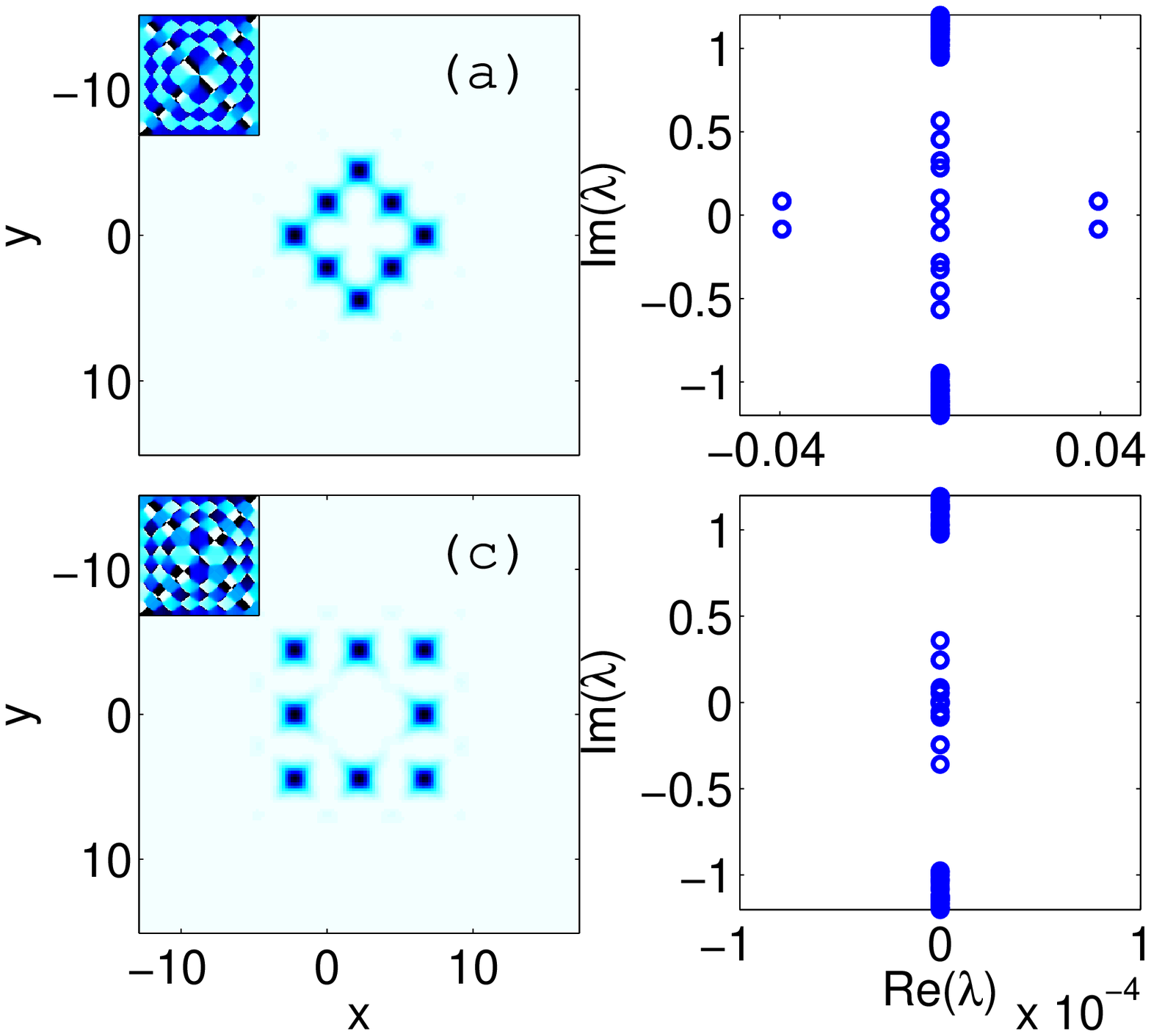}
\includegraphics[width=8cm,height=7cm,angle=0,clip]{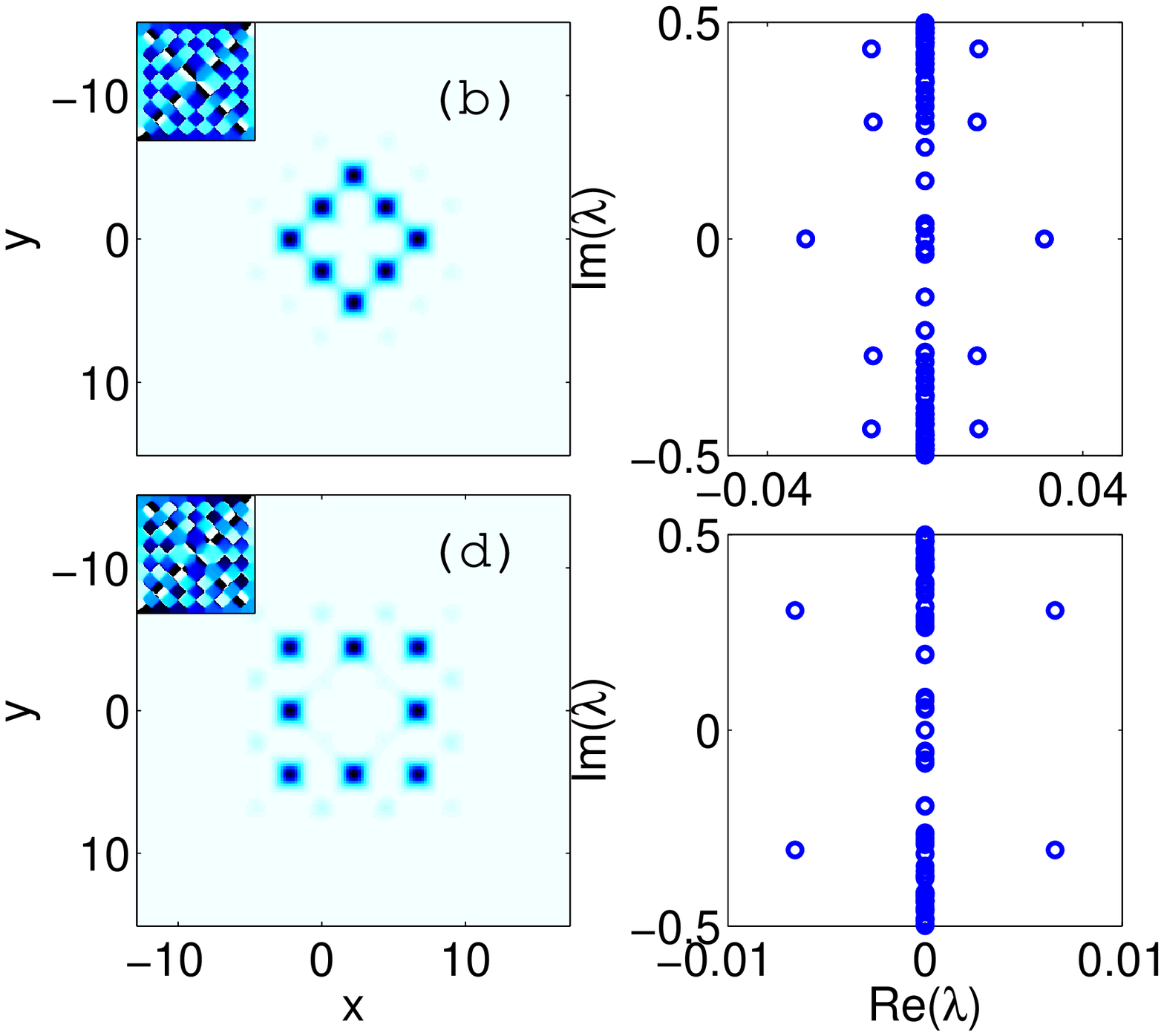}
\caption{(Color online) Density Profile (with phase inset)
and linearization spectrum for the case of a
defocusing nonlinearity for $\beta=9.2$ (a,c) and $9.9$ (b,d).
The latter is very close to the band edge which in this case
is at $\beta \approx 10$.}
\label{cont_4}
\end{figure}

\subsection{Defocusing case}

For completeness, we consider the defocusing case briefly as well.
The theoretical predictions for the discrete defocusing model
are tantamount to the ones for the focusing case because the staggering
transformation $\tilde{v}_{m,n} = (-1)^{m+n} v_{m,n}$, which takes a
solution $v$ of the focusing problem to a solution $\tilde{v}$ of the
defocusing problem,
does not affect
next-nearest-neighbor configurations and takes the nearest-neighbor
8-site $S=2$ vortex to an equivalent $S=-2$ (similarly to what
happens for the nearest neighbor 4-site $S=1$ vortex).

In the continuum version of the model with the saturable nonlinearity,
we use the transformation $\tilde{E}=-E$.
The configurations now live shifted by one half period of the lattice
to the right (i.e. each letter is shifted to the nearest minima to
the right in the bottom right panel of Fig. \ref{cont_1}).
The linear spectrum shifts and the localized solutions are now
found within the first band-gap (as opposed to the semi-infinite gap for
the focusing case).
Similarly to the discrete model, again the principal
predictions persist, as seen in Fig. \ref{cont_4}.
I.e., there is a real pair of eigenvalues
close to the band-edge for the square configuration and again
there is a large stability region for the rhomboidal one.
One difference is that the square configuration is now always unstable
in the entire first bandgap due to complex quartets of eigenvalues
(see Fig. \ref{cont_4}(a) as an example). An example
of this configuration after the real pair bifurcates through the
origin is given in Fig. \ref{cont_4}(b).
Rhomboidal solutions from before (c) and after (d) the
destabilization close to the band edge
are also shown in
Fig. \ref{cont_4} along with their linearization spectra.  These solutions
again disappear near to the band edge
(via saddle-node bifurcations) and do not bifurcate from
linear modes.

\begin{figure}[t]
\includegraphics[width=18cm,height=8cm,angle=0,clip]{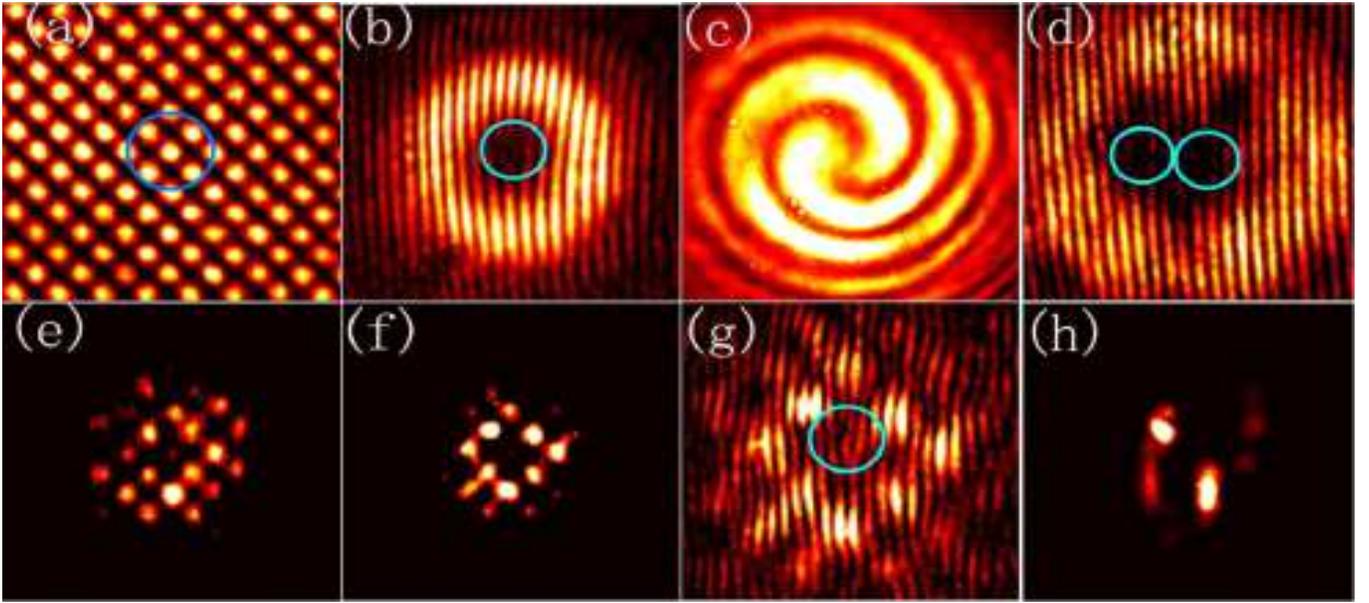}
\caption{(Color online) Experimental observation of a double-charge
discrete vortex
soliton extended to eight lattice sites under self-focusing nonlinearity.
Top panels: (a) Input
lattice beam
pattern where the circle indicates the location of the
S=2 vortex beam at input,
(b, c) interference patterns of the input vortex with an inclined
 plane wave (b) and with a spherical wave (c), (d) output of the S=2
 vortex after $10$ mm of linear propagation through the crystal,
 showing
the breakup into two well-separated S=1 vortices.
Bottom panels: (e) output vortex pattern at a low bias field,
(f) self-trapped S=2 vortex pattern at a high bias field,
(g) interference pattern of  the vortex soliton with an inclined plane
wave (zoomed), and (h) nonlinear output of the double-charge vortex
without
the lattice. Small circles in the interferograms mark the locations of
vortex singularities.}
\label{fig6}
\end{figure}

\begin{figure}[t]
\includegraphics[width=18cm,height=4cm,angle=0,clip]{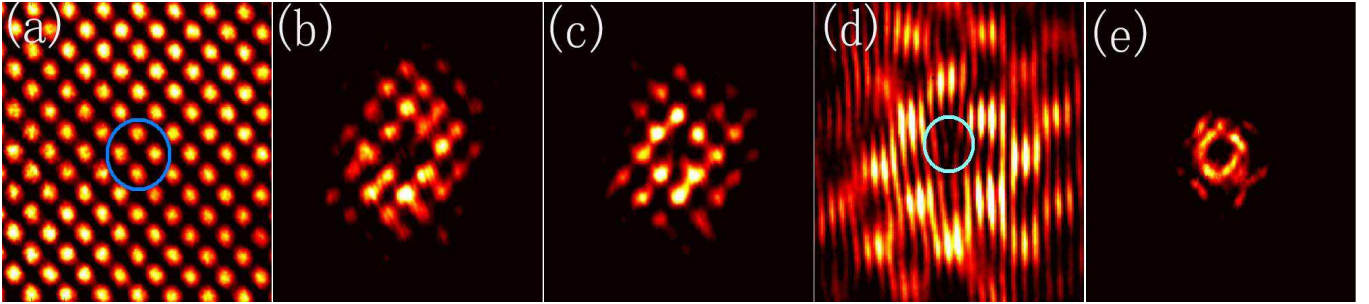}
\caption{(Color online) Experimental observation of an extended S=2
gap vortex soliton in photonic lattices with self-defocusing nonlinearity.
(a) Similar to Fig. \ref{fig6}(a) but the vortex location is different,
(b) output vortex pattern at a low bias field (-700V/cm), (c) self-trapped
vortex pattern at a high bias field (-1.5kV/cm),
(d) zoom-in interferogram as in Fig. 7(g), and
(e) the Fourier-space spectrum of the S=2 vortex soliton.}
\label{fig7}
\end{figure}

\section{Experimental Results}

In our experiment, we use a setup similar to that used in
\cite{vortex1}
for the observation of fundamental ($S=1$) discrete vortex solitons.
The square lattice is induced in a biased photorefractive crystal
(SBN:60 $5 \times 10 \times 5$ mm$^3$) by a spatially
modulated partially coherent laser
beam ($\lambda=488$ nm) sent through an amplitude mask. When the mask
is
appropriately imaged onto the input face of the crystal, the
 Talbot effect of the periodically modulated laser beam is suppressed
by using a diffraction element, so the lattice intensity pattern
remains
invariant during the propagation throughout the crystal. The
double-charged (S=2) vortex beam is generated by sending a coherent
laser beam through a computer generated vortex hologram. In the
experiment,
the lattice beam is ordinarily-polarized and the vortex beam is
extraordinarily-polarized. Thus the lattice beam will undergo nearly
linear propagation while the vortex beam will experience a large
nonlinearity due to the anisotropic photorefractive property.
The input/output intensity patterns of the vortex beams are monitored
with
 CCD cameras. In addition, the vortex beam exiting the crystal is also
sent into a Mach-Zehnder interferometer for phase measurement, as
needed.

To observe self-trapping of doubly-charged vortex solitons, the
donut-like vortex beam is expended and launched into the lattice such
that the vortex ring covers eight lattice sites
[indicated by a blue  circle in Fig. \ref{fig6}(a)], while the vortex
core is
overlapping with the central non-excited site. This arrangement
corresponds
to the square vortex configuration (ABCDEFG) illustrated in
Fig. \ref{cont_1}.
The phase singularity of the input vortex is identified from two
different interferograms shown in Figs. \ref{fig6}(b, c)
(zoomed in so as to see the fringes more clearly).
Two-fork fringes [Fig. \ref{fig6}(b)] and two-start spirals
[Fig. \ref{fig6}(c)] in the central region of the interferograms
clearly show the S=2 phase dislocations. As expected, the $S=2$ vortex
breaks  up into two S=1 vortices during linear propagation through the
homogenous medium (i.e., without the waveguide lattice), as shown in
Fig.
\ref{fig6}(d). This is the natural outcome of
the topological instability \cite{beks}.
When a dc field is applied along the crystalline c-axis, the SBN
crystal
turns into a self-focusing medium
\cite{solit,moti1,moti2,dip,quad,neck}.
Under a proper strength of
the nonlinearity, the vortex beam evolves into a $S=2$ vortex soliton.
 Typical results are shown in the bottom panels of Fig. \ref{fig6},
for which the lattice period is about $25 \mu$m and the
vortex-to-lattice intensity ratio of the beams is about $1:4$.
At a low bias field of $800$V/cm, the output vortex exhibits typical
discrete diffraction covering many lattice sites [Fig. \ref{fig6}(e)].
However, when the bias field is increased to about $1.6$kV/cm,
the vortex beam self-traps into a $S=2$ vortex soliton, covering
mainly the
eight sites excited at the input [Fig. \ref{fig6}(f)]. In order to
identify
the phase structure of the nonlinear localized state, a tilted broad
beam
(akin to a quasi-plane wave) is sent to interfere with the vortex
soliton at
the output. It can be seen clearly [Fig. \ref{fig6}(g)] that
the two
forks remain in the center, and their bifurcation directions remain
also
unchanged as compared with the input vortex beam [Fig. \ref{fig6}(b)].
 This provides a direct evidence for the formation of $S=2$
high-order discrete vortex solitons with preserving phase
singularities. We emphasize that such solitons are generated under the
8-site excitation, as predicted in theory, and they are quite
different from the 4-site
excitation which leads to dynamical charge flipping or disintegration
of the topological charge, as observed in our previous experiment
\cite{zc_s2}. Without the lattice, the $S=2$ vortex becomes unstable
and
breaks
up into soliton filaments due to azimuthal modulation instability
under the same bias conditions [Fig. \ref{fig6}(h)].  This indicates
that, in an optically induced square lattice, the S=2 vortex can be
geometrically stabilized and self-trapped, in good agreement with above
theoretical and numerical results.

Finally, Fig. \ref{fig7} shows our experimental results on self-trapping of extended S=2 vortex in the photonic square lattices induced with a self-defocusing nonlinearity.  Notice that the location of the vortex ring [illustrated by the blue circle in Fig. \ref{fig7}(a)] is different from that in Fig. \ref{fig6}(a) in order to have the extended 8-site excitation, now that the waveguides are located at the intensity minima (rather than maxima)
of the lattice-inducing beam under self-defocusing nonlinearity \cite{dsong}.
In comparison with very recent study of S=2 discrete vortex solitons in
hexagonal photonic lattices \cite{ourhexhon,ournewpra},
we found that the S=2 vortex can remain self-trapped [Fig. \ref{fig7}(c)] and
maintain its singularities [Fig. \ref{fig7}(d)] also under the self-defocusing
nonlinearity. In addition, differently from the more localized
4-site excitation
of the S=2 vortex which evolves into a self-trapped quadrupole-like structure
\cite{dsong}, here the vortex phase structure remains. Furthermore, the
Fourier-space spectrum [Fig. \ref{fig7}(e)] does not concentrate near the
four  M-points of the first Broullion zone, confirming the
above numerical finding that the complex mode
structures of such high-order vortex solitons do not bifurcate
from the edge of the Bloch band.

\section{Conclusions and Future Challenges}

In the present work, we have studied
the geometric stabilization of S=2 vortices through the presence of next-nearest neighbor interactions in 2D square lattices. Different orientations for multi-site excitation of the vortices were studied and compared with previous work where the geometric stabilization is absent and nearest-neighbor interactions mediated through
the central site leading to instability. Based on the results obtained from the standard cubic nonlinear
Schr{\"o}dinger lattice model, we extended our study
to the case of the discrete model with a saturable nonlinearity, revealing
some subtleties with respect to next-nearest neighbor
eigenvalue calculations. Nevertheless, the principal analytical
and numerical observations persisted therein. These findings
were further studied by computations in the full continuum model, and we found that both the square and rhomboidal orientations of the extended vortex can lead to stable S=2 vortex solitons under appropriate conditions. Lastly, we
corroborated these analytical and numerical results with
experimental observations in optically induced photonic lattices.
In a 10 mm long photorefractive crystal, we demonstrated
that a double-charge vortex can maintain its singularity
during nonlinear propagation in square lattices under
both self-focusing and -defocusing nonlinearities.

It would be interesting to experimentally study the dynamics of
higher charge vortices,
such as vortices of topological charge $S=3$, for which the discrete model
theory predicts
potential stability, as well as to consider using crystals with longer propagation distances such that some of the
above-predicted instability phenomena could be observable. On the other hand, from a theoretical perspective, a detailed understanding of vortices in more complex lattice settings
such as superlattices and quasi-crystalline lattices
could be important for relevant studies in other nonlinear systems involving vortices and periodic potentials.

\begin{acknowledgments}
This work was supported by the 973 program, NSFC, PCSIRT, NSF,
AFOSR and the Alexander von Humboldt Foundation.
We thank C. Lou, L. Tang and J. Yang for discussions and assistance.

\end{acknowledgments}

\end{document}